\documentclass[twocolappendix,appendixfloats]{emulateapj}

\usepackage{apjfonts}
\usepackage{color}
\usepackage{CJK}

\newcommand{\pivec}{\mbox{\boldmath $\pi$}}
\newcommand{\muvec}{\mbox{\boldmath $\mu$}}
\newcommand{\te}{t_{\rm E}}
\newcommand{\thetae}{\theta_{\rm E}}
\newcommand{\pie}{\pi_{\rm E}}

\def\e{{\rm E}}

\definecolor{darkbrown}{RGB}{139,69,19}

\shorttitle{Super-Earth Orbiting a Low-mass Star}
\shortauthors{HAN ET AL.}

\begin{document}

\title{ OGLE-2017-BLG-0482L\lowercase{b}: A Microlensing Super-Earth Orbiting a Low-mass Host Star}

\author{
C.~Han\altaffilmark{001}, 
Y.~Hirao\altaffilmark{002,102},
A.~Udalski\altaffilmark{003,101}, 
C.-U.~Lee\altaffilmark{004,103},
V.~Bozza\altaffilmark{005,006}, 
A.~Gould\altaffilmark{004,007,008,103},\\
and\\
F.~Abe\altaffilmark{009}, R.~Barry\altaffilmark{010}, I.~A.~Bond\altaffilmark{011}, 
D.~P.~Bennett\altaffilmark{010,012}, A.~Bhattacharya\altaffilmark{010,012}, 
M.~Donachie\altaffilmark{013}, P.~Evans\altaffilmark{013}, A.~Fukui\altaffilmark{014}, 
Y.~Itow\altaffilmark{009}, K.~Kawasaki\altaffilmark{002}, 
N.~Koshimoto\altaffilmark{002}, M.~C.~A.~Li\altaffilmark{013}, C.~H.~Ling\altaffilmark{011}, 
Y.~Matsubara\altaffilmark{009}, S.~Miyazaki\altaffilmark{002}, 
H.~Munakata\altaffilmark{009}, Y.~Muraki\altaffilmark{009}, M.~Nagakane\altaffilmark{002}, 
K.~Ohnishi\altaffilmark{015}, C.~Ranc\altaffilmark{010}, N.~Rattenbury\altaffilmark{013}, 
T.~Saito\altaffilmark{016}, A.~Sharan\altaffilmark{013}, D.~J.~Sullivan\altaffilmark{017}, 
T.~Sumi\altaffilmark{002}, D.~Suzuki\altaffilmark{018}, P.~J.~Tristram\altaffilmark{019},
T.~Yamada\altaffilmark{002}, A.~Yonehara\altaffilmark{020}\\
(The MOA Collaboration),\\
P.~Mr\'oz\altaffilmark{003}, R.~Poleski\altaffilmark{003,007}, S.~Koz{\l}owski\altaffilmark{003},
I.~Soszy\'nski\altaffilmark{003}, P.~Pietrukowicz\altaffilmark{003}, J.~Skowron\altaffilmark{003}, 
M.~K.~Szyma\'nski\altaffilmark{003}, K.~Ulaczyk\altaffilmark{003}, M.~Pawlak\altaffilmark{003},
K.~Rybicki\altaffilmark{003}, P.~Iwanek\altaffilmark{003}\\
(The OGLE Collaboration) \\   
M.~D.~Albrow\altaffilmark{021}, S.-J.~Chung\altaffilmark{004,022}, K.-H.~Hwang\altaffilmark{004}, 
Y.~K.~Jung\altaffilmark{023}, D.~Kim\altaffilmark{001}, W.-T.~Kim\altaffilmark{024}, 
H.-W.~Kim\altaffilmark{004}, Y.-H.~Ryu\altaffilmark{004}, I.-G.~Shin\altaffilmark{023}, 
Y.~Shvartzvald\altaffilmark{027,104}, J.~C.~Yee\altaffilmark{023}, W.~Zhu\altaffilmark{025},       
S.-M.~Cha\altaffilmark{004,026}, S.-L.~Kim\altaffilmark{004,022}, D.-J.~Kim\altaffilmark{004}, 
D.-J.~Lee\altaffilmark{004}, Y.~Lee\altaffilmark{004,026},
B.-G.~Park\altaffilmark{004,022}, R.~W.~Pogge\altaffilmark{007}   \\ 
(The KMTNet Collaboration),\\
}

\email{cheongho@astroph.chungbuk.ac.kr}

\altaffiltext{001}{Department of Physics, Chungbuk National University, Cheongju 28644, Republic of Korea} 
\altaffiltext{002}{Department of Earth and Space Science, Graduate School of Science, Osaka University, Toyonaka, Osaka 560-0043, Japan}
\altaffiltext{003}{Warsaw University Observatory, Al. Ujazdowskie 4, 00-478 Warszawa, Poland} 
\altaffiltext{004}{Korea Astronomy and Space Science Institute, Daejon 34055, Republic of Korea} 
\altaffiltext{005}{Dipartimento di Fisica "E. R. Caianiello", Universit\'a di Salerno, Via Giovanni Paolo II, I-84084 Fisciano (SA), Italy} 
\altaffiltext{006}{Istituto Nazionale di Fisica Nucleare, Sezione di Napoli, Via Cintia, I-80126 Napoli, Italy} 
\altaffiltext{007}{Department of Astronomy, Ohio State University, 140 W.\ 18th Ave., Columbus, OH 43210, USA} 
\altaffiltext{008}{Max Planck Institute for Astronomy, K\"onigstuhl 17, D-69117 Heidelberg, Germany}
\altaffiltext{009}{Institute for Space-Earth Environmental Research, Nagoya University, 464-8601 Nagoya, Japan} 
\altaffiltext{010}{Code 667, NASA Goddard Space Flight Center, Greenbelt, MD 20771, USA} 
\altaffiltext{011}{Institute of Natural and Mathematical Sciences, Massey University, Auckland 0745, New Zealand} 
\altaffiltext{012}{Deptartment of Physics, University of Notre Dame, 225 Nieuwland Science Hall, Notre Dame, IN 46556, USA} 
\altaffiltext{013}{Dept.~of Physics, University of Auckland, Private Bag 92019, Auckland, New Zealand} 
\altaffiltext{014}{Okayama Astrophysical Observatory, National Astronomical Observatory of Japan, Asakuchi,719-0232 Okayama, Japan} 
\altaffiltext{015}{Nagano National College of Technology, 381-8550 Nagano, Japan} 
\altaffiltext{016}{Tokyo Metroplitan College of Industrial Technology, 116-8523 Tokyo, Japan} 
\altaffiltext{017}{School of Chemical and Physical Sciences, Victoria University, Wellington, New Zealand} 
\altaffiltext{018}{Institute of Space and Astronautical Science, Japan Aerospace Exploration Agency, Kanagawa 252-5210, Japan}
\altaffiltext{019}{Mt.~John University Observatory, P.O. Box 56, Lake Tekapo 8770, New Zealand} 
\altaffiltext{020}{Department of Physics, Faculty of Science, Kyoto Sangyo University, 603-8555 Kyoto, Japan}
\altaffiltext{021}{University of Canterbury, Department of Physics and Astronomy, Private Bag 4800, Christchurch 8020, New Zealand}
\altaffiltext{022}{Korea University of Science and Technology, 217 Gajeong-ro, Yuseong-gu, Daejeon 34113, Republic of Korea}
\altaffiltext{023}{Harvard-Smithsonian Center for Astrophysics, 60 Garden St., Cambridge, MA, 02138, USA}
\altaffiltext{024}{Department of Physics \& Astronomy, Seoul National University, Seoul 151-742, Republic of Korea} 
\altaffiltext{025}{Canadian Institute for Theoretical Astrophysics, University of Toronto, 60 St George Street, Toronto, ON M5S 3H8, Canada}
\altaffiltext{026}{School of Space Research, Kyung Hee University, Yongin 17104, Republic of Korea}
\altaffiltext{027} {Jet Propulsion Laboratory, California Institute of Technology, 4800 Oak Grove Drive, Pasadena, CA 91109, USA} 
\altaffiltext{101}{OGLE Collaboration}
\altaffiltext{102}{The MOA Collaboration}
\altaffiltext{103}{The KMTNet Collaboration}
\altaffiltext{104}{NASA Postdoctoral Program Fellow}

\begin{abstract}
We report the discovery of a planetary system in which a super-earth orbits a 
late M-dwarf host.  The planetary system was found from the analysis of the 
microlensing event OGLE-2017-BLG-0482, wherein the planet signal appears as 
a short-term anomaly to the smooth lensing light curve produced by the host.  
Despite its weak signal and short duration, the planetary signal was firmly 
detected from the dense and continuous coverage by three microlensing surveys.   
We find a planet/host mass ratio of $q\sim 1.4\times 10^{-4}$.  We measure 
the microlens parallax $\pie$ from the long-term deviation in the observed 
lensing light curve, but the angular Einstein radius $\thetae$ cannot be 
measured because the source trajectory did not cross the planet-induced 
caustic. Using the measured event timescale and the microlens parallax, we 
find that the masses of the planet and the host are 
$M_{\rm p}=9.0_{-4.5}^{+9.0}\ M_\oplus$ and 
$M_{\rm host}=0.20_{-0.10}^{+0.20}\ M_\odot$, respectively, and the projected 
separation between them is $a_\perp=1.8_{-0.7}^{+0.6}$~au.  The estimated 
distance to the lens is $D_{\rm L}=5.8_{-2.1}^{+1.8}$ kpc.  The discovery 
of the planetary system demonstrates that microlensing provides an important 
method to detect low-mass planets orbiting low-mass stars.  
\end{abstract}


\section{Introduction}

\begin{figure*}
\epsscale{0.75}
\plotone{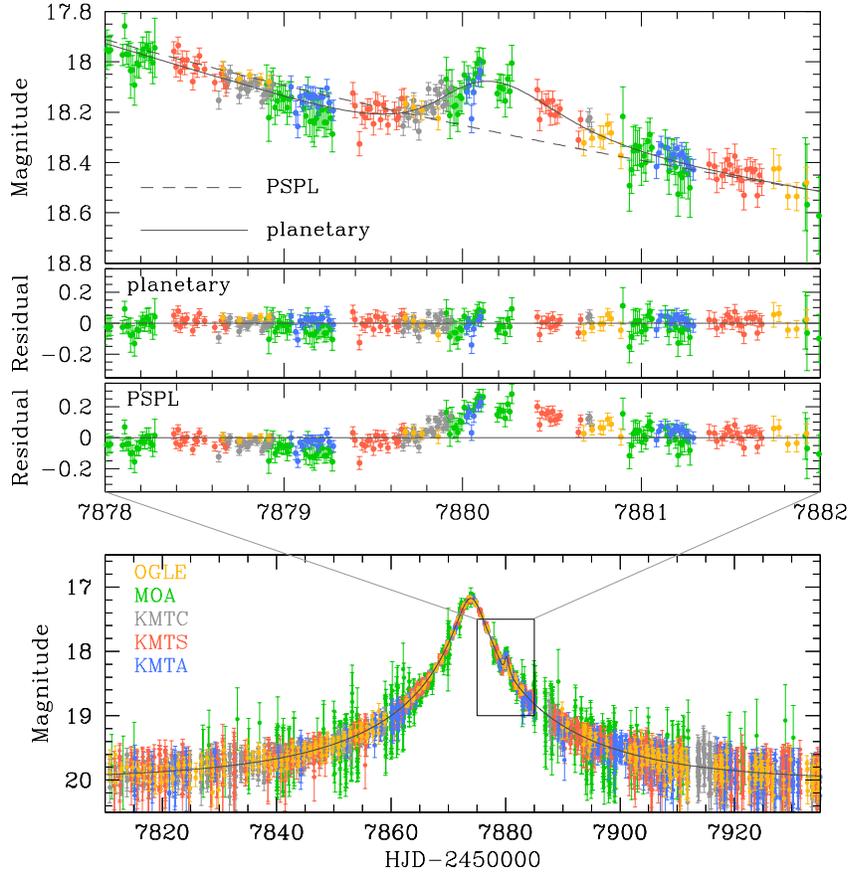}
\caption{
Light curve of the gravitational microlensing event OGLE-2017-BLG-0482. The 
upper panels shows the zoom of the planetary anomaly region that is enclosed by 
a box in the bottom panel. The solid and dashed curves superposed on the data 
are the best-fit planetary model and the point-source point-lens (PSPL) model, 
respectively.  The middle two panels show the residuals from the individual models. 
}
\label{fig:one}
\end{figure*}

The last two decades have witnessed a great increase of the number of known 
planetary systems, which is reaching $\sim 2800$ as of the end of 2017 according 
to the Extrasolar Planet Encyclopedia (http://exoplanet.eu/catalog/). Notably, 
the {\it Kepler} mission using the transit method has contributed to the dramatic 
increase in the number of known planetary systems. However, most of the known planets 
belong to solar-type stars, and the number of planets orbiting low-mass stars is much 
smaller than that of planets orbiting solar-type stars. In particular, planets around 
late M-type dwarfs comprise a very small portion of known planets due to the difficulty 
of observing their host stars.

Planets around low-mass stars may be formed through a different process 
from those orbiting solar-type stars.  For example, the luminosity of 
low-mass stars decreases substantially (by a factor of $\sim 10$ -- 100) on the 
Hayashi track during typical planet formation timescales of 10 -- 100 Myrs, while 
the luminosity of solar-mass stars remains almost constant \citep{Kennedy2006}. 
The difference in the environment between the two types of stars would likely 
affect the evolution of stellar disks and the subsequent planet formation processes.  
See the review of \citet{Boss1989}  about the planet formation process of low-mass 
stars.  However, the details of the planet formation process for low-mass stars 
are poorly known because the planet sample is too small to check proposed scenarios.  
As a result, our understanding of planets around low-mass stars is incomplete despite 
the fact that the hosts are the most common population of stars in the Galaxy.

Microlensing occurs by the gravitational field of an intermediary objects between 
an observer and a background star.  Due to this nature, planet detections using 
the microlensing method do not rely on the host's luminosity but just its gravity 
and that of the planet, while other planet detection methods rely on the luminosity 
of the host.  This enables one to extend microlensing planet searches to stars with 
very low luminosities and even substellar brown-dwarf hosts, e.g.\ OGLE-2012-BLG-0358Lb 
\citep{Han2013}.  Furthermore, the method is sensitive to low-mass planets down to 
Earth-mass planets, e.g., OGLE-2016-BLG-1195Lb \citep{Shvartzvald2017, Bond2017}.  
For this reason, microlensing planets comprise $\sim 23\%$ of the known planets with 
host masses $\lesssim 0.2~M_\odot$ and planet masses $\lesssim 10~M_\oplus$, although 
they comprise only $\sim 2\%$ of the total planet sample.

In this work, we report the microlensing discovery of a super-earth planet orbiting 
a low-mass M-dwarf host.  The planetary system was found from the analysis of the 
microlensing event OGLE-2017-BLG-0482, in which the planet revealed its presence 
as a short-term anomaly. Despite the short duration, the planet signal 
was firmly detected from the combination of three high-cadence lensing surveys.

\section{Observations and Data}
 
In Figure~\ref{fig:one}, we present the light curve of the lensing event 
OGLE-2017-BLG-0482. The  event occurred on a faint star with a baseline 
magnitude $I\sim 20$. The source star is located toward the Galactic bulge 
field  with equatorial coordinates 
$({\rm RA},{\rm DEC})_{\rm J2000}=$ (17:56:11.73, -30:31:42.1), which 
correspond to the Galactic coordinates $(l,b)=(-0.2^\circ, -2.8^\circ)$.  
The amplification of the source flux induced by lensing was first noticed 
on April 8, 2017 (${\rm HJD}'={\rm HJD} - 2450000\sim 7852$) by the Optical 
Gravitational Lensing Experiment \citep[OGLE:][]{Udalski2015} survey that 
is conducted using the 1.3m Warsaw telescope at Las Campanas Observatory 
in Chile. Images of the OGLE survey were taken mainly in $I$ band, and some 
$V$-band images were taken for color measurement.

The event was also in the observation fields of the Microlensing Observations 
in Astrophysics \citep[MOA:][]{Bond2001, Sumi2003} survey and the Korea 
Microlensing Telescope Network \citep[KMTNet:][]{Kim2016} survey. Data of 
the MOA survey were taken using the 1.8m telescope located at the Mt.~John 
University Observatory in New Zealand. 
In the list of MOA transient events, the event is denoted 
by MOA-2017-BLG-209. MOA data were acquired using a customized $R$ band that 
has a bandwidth corresponding to roughly the sum of the standard $R$ and $I$ 
bands.
The KMTNet data were obtained using 
its three globally distributed 1.6m telescopes located at the Cerro Tololo 
Interamerican Observatory in Chile (KMTC), the South African Astronomical 
Observatory in South Africa (KMTS), and the Siding Spring Observatory in 
Australia (KMTA).  KMTNet observations were conducted in $I$- and $V$-band filters. 
The KMTNet data obtained by each telescope are composed of two sets (denoted by 
BLG01 and BLG41) because the survey alternately covered the field with a 
$6^\prime$ offset to fill gaps between the camera chips.  In the list of 
KMTNet microlensing candidates, it is called SAO01M0605.043904 
\citep{KimDJ2018,KimHW2018}.

After it was detected, the light curve of the event followed the smooth form of 
a single-mass lensing event reaching a peak magnification $A_{\rm max}\sim 16$ 
at ${\rm HJD}'\sim 7874$.  On May 6 UT 15:29 (${\rm HJD}'\sim 7880.15$), the 
MOA group alerted the microlensing community to a possible planetary anomaly 
based on real-time assessment by the MOA observer with the aim of encouraging 
follow-up observations.   Unfortunately, no follow-up observation could be 
conducted mainly due to the short duration of the anomaly.  A day after the 
anomaly alert, Y.~Hirao of the MOA group released a model of the anomaly 
based on the MOA data.  According to this model, the anomaly was produced 
by the crossing of the source over the caustic produced by a planetary 
companion with a mass ratio of $q\sim  10^{-4}$.  V.~Bozza also released a 
similar model.  From modeling conducted with the addition 
of data from the OGLE and KMTNet surveys, it was noticed that the earlier 
models exhibit inconsistency with the additional data, and an updated model 
without caustic crossing was presented by C.~Han.  After the anomaly, the 
event followed the light curve of a single-mass event and gradually returned 
to the baseline.

The firm detection and characterization of this weak and short planetary 
signal was made possible by the combination of the three high-cadence 
lensing surveys.  In the upper panel of Figure~\ref{fig:one}, we present 
the zoom of the planet-induced anomaly.  The anomaly lasted only for about 
2 days.  Furthermore, the signal is weak with a maximum deviation of 
$\sim 0.2$ mag relative to the single-mass lensing light curve.  Nevertheless, 
the signal was densely and continuously covered by the survey experiments 
thanks to the high-cadence observations conducted using globally distributed 
telescopes.

Photometry of the data is conducted using softwares customized by the individual 
groups based on the Difference Imaging Analysis \citep{Alard1998, Wozniak2000}: 
\citet{Udalski2003} for the OGLE, \citet{Bond2001} for the MOA groups, and 
pyDIA developed by M.~Albrow for the KMTNet.  For the KMTC data set, an additional 
photometry is conducted with DoPHOT software \citep{Schechter1993} for the 
determination of the source color and the construction of a color-magnitude 
diagram.  For the use of heterogeneous data sets obtained using different 
instruments and processed using different photometry codes,  error bars of 
the individual data sets are readjusted following the procedure described 
by \citet{Yee2012}.

\begin{figure}
\includegraphics[width=\columnwidth]{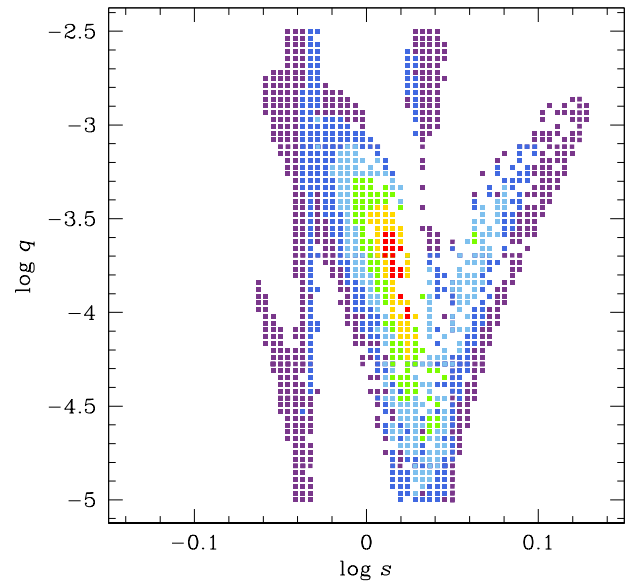}
\caption{
Distribution of $\Delta\chi^2$ in the $(\log s, \log q)$ plane.  
Locations marked in different colors represent the regions with 
$\Delta\chi^2 < n^2$ (red), $(2n)^2$ (yellow), $(3n)^2$ (green), $(4n)^2$ 
(cyan), $(5n)^2$ (blue), and $(6n)^2$ (purple) from the best-fit solution, 
where $n=5$.  }
\label{fig:two}
\end{figure}

\section{Analysis}

\subsection{Planetary Analysis}

The observed light curve appears to be a typical case of a planetary lensing 
event in which the planetary signal is revealed as a short-term perturbation to 
the smooth light curve produced by the host of the planet \citep{Mao1991, 
GouldLoeb1992}. For the basic description of the lensing light curve produced 
by a lens composed of two masses, one requires six parameters.  Three of these 
parameters describe the geometry of the lens-source approach: the time of 
the closest lens-source separation, $t_0$, the lens-source separation at 
that time, $u_0$ (normalized to the angular Einstein radius $\thetae$), 
and the time for the source to cross the Einstein radius, $t_{\rm E}$ 
(Einstein timescale). The other three parameters describe the binarity of the 
lens: the separation between the primary ($M_1$) and the companion ($M_2$), 
$s$ (also normalized to $\thetae$), their mass ratio, $q=M_2/M_1$, and the 
angle between the source trajectory and the binary axis, $\alpha$ (source 
trajectory angle). If the source crosses the planet-induced caustic, the 
perturbation is affected by finite-source effects \citep{Bennett1996}.  
In such cases, one needs an additional parameter $\rho=\theta_*/\thetae$ 
(normalized source radius) to account for the light curve deviation caused 
by the finite-source effect. Here $\theta_*$ represents the angular radius 
of the source star.

\begin{figure}
\includegraphics[width=\columnwidth]{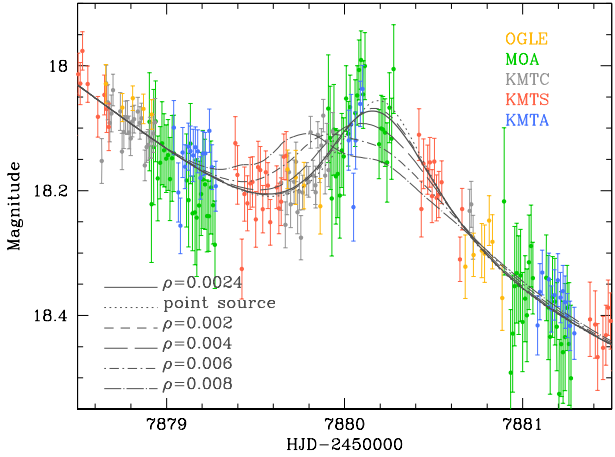}
\caption{
Model light curves expected from various values of the normalized source 
radius $\rho$.
}
\label{fig:three}
\end{figure}

In cases for which planetary signals can be treated as perturbations, one 
can heuristically estimate the lensing parameters related to the planet, 
i.e., $s$ and $q$, from the location and duration of the perturbation 
\citep{GouldLoeb1992, Gaudi2012}. A planet induces two types of caustics: 
central and planetary caustics. The central caustic lies close to the 
primary of the lens, and thus the caustic produces perturbations near the 
peak of high magnification events. The planetary caustic, on the other 
hand, lies away from the primary lens with a separation 
\begin{equation} u_{\rm p}=s - {1 \over s} , \end{equation} and thus 
perturbations induced by the planetary caustic can appear anywhere along 
the lensing light curve. The perturbation of OGLE-2017-BLG-0482 lies away 
from the peak and thus it is produced by a planetary caustic. The 
perturbation occurred when the lensing magnification was 
$A=(u^2+2)/[u(u^2+4)^{1/2}]\sim 7.5$, i.e., when the lens-source separation 
was $u\sim 0.135$.  By substituting $u$ into $u_{\rm p}$ in Equation (1), 
one finds that the planet-host separation is either $s\sim 0.93$ or 1.07, 
which are referred to as the close ($s<1$) and wide-separation ($s>1$) 
solutions, respectively.  The caustic induced by a close planet usually 
produces a dip (negative deviation) in the light curve, while the caustic 
induced by a wide planet always produces a bump (positive deviation).  
The observed bump structure of the perturbation, therefore, suggests 
that the planet separation is greater than $\thetae$, i.e., $s\sim 1.07$.

The duration of the planet-induced perturbation results from the combination 
of the sizes of the caustic, $\Delta\xi_{\rm p}$, and the source, $\rho$.  
If the source is bigger than the caustic, $\rho > \Delta\xi_{\rm p}$, the 
duration corresponds to the source crossing time, 
$\Delta t_{\rm p} \sim 2\rho t_{\rm E}$, and thus is mostly determined by 
the source radius.  If the source is smaller than the caustic, on the 
other hand, the duration is determined by the size of the caustic. In 
the case of OGLE-2017-BLG-0482, the source is a very faint main-sequence 
star and thus the duration is likely to depend on the caustic size. The 
size of the planetary caustic is related to the separation and the mass 
ratio between the planet and the host by \citep{Han2006}
\begin{equation}
\Delta\xi_{\rm p} = {4q^{1/2} \over s^2} \left( 1+{1\over 2s^2} \right).
\end{equation}
Since the duration of the planetary signal is 
$\Delta t_{\rm p}\sim \Delta\xi_{\rm p}t_{\rm E}$ , the mass ratio is expressed by
\begin{equation}
q = \left( {s^4 \over 4s^2+2} {\Delta t_{\rm p} \over t_{\rm E}} \right)^2 . 
\end{equation}
With $s\sim 1.07$, $\Delta t_{\rm p}\sim 2$ days, and $t_{\rm E}\sim 40$ days, one 
finds that the mass ratio is $\sim 10^{-4}$.

For the accurate determinations of the lensing parameters, we conduct 
numerical modeling of the observed light curve. We search for the solution 
of the lensing parameters in two steps.  In the first step, we conduct a 
dense grid search over $(\log s, \log q)$ plane.  At each point on this 
plane, we hold the two grid parameters fixed while allowing the remaining 
five parameters $(t_0,u_0,\te,\rho, \alpha)$ to vary in six Markov Chain 
Monte Carlo (MCMC) $\chi^2$ minimizations that are equally spaced around the circle 
in their seed values of $\alpha$. The seed values of $(t_0,u_0,\te)$ are 
taken from the point-lens fit, and we seed $\rho$ at $\rho=1.0\times 10^{-3}$.  
From this preliminary search, we identify local $\chi^2$ minima on the 
$(\log s, \log q)$ plane.  In the second step, we refine the individual 
local minima by allowing all parameters to vary. Figure~\ref{fig:two} 
displays the $\Delta\chi^2$ map over the $(\log s,\log q)$ plane obtained 
from the grid search. It shows that there exists a unique planetary 
solution with the planet separation slightly greater than unity and 
a mass ratio of $q=$ (1 -- 3) $\times 10^{-4}$, which roughly matches 
the prediction of the heuristic analysis.

\begin{figure}
\includegraphics[width=\columnwidth]{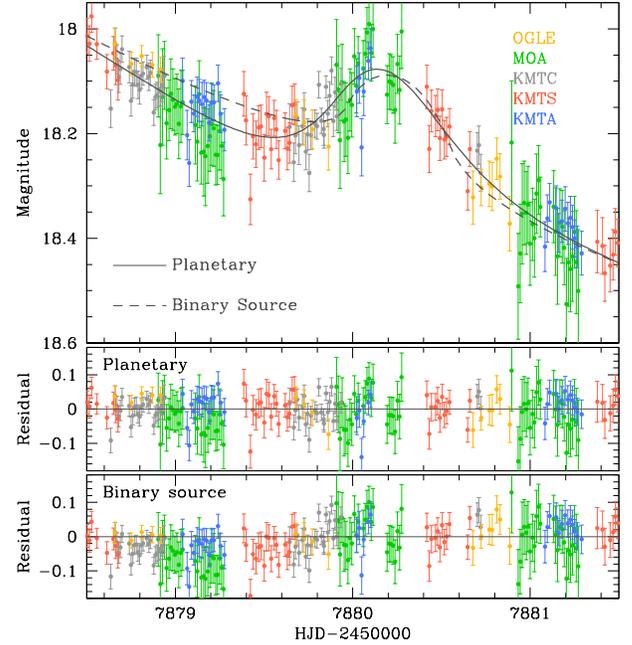}
\caption{
Comparison of the planetary (solid curve) and binary-source (dashed curve) 
models.  The two lower panels show the residuals from the individual models. 
}
\label{fig:four}
\end{figure}

\subsection{Finite-source Effects}

When a source passes over or approaches very close to caustics, the 
planetary anomaly is affected by finite-source effects and the analysis 
of the deviation enables one to measure the normalized source radius 
$\rho$.  
In the case of OGLE-2017-BLG-0482, it is found that $\rho$ cannot be measured 
due to the lack of finite-source effects. Measuring $\rho$ is important because 
the angular Einstein radius, which is needed to determine the lens mass, is 
estimated from $\rho$ by $\thetae=\theta_*/\rho$. The difficulty of the $\rho$ 
measurement is caused by the fact that the source did not cross the caustic.

Nevertheless, the source approached close to a strong cusp of the 
caustic around which the gradient of lensing magnification is high. In 
this case, one can set an upper limit on the source size.  To test this 
possibility, we draw light curves expected from various values of $\rho$.  
See Figure~\ref{fig:three}. From this, combined with the MCMC chain 
obtained from modeling, we find that the $3\,\sigma$ upper 
limit of the normalized source size is $\rho_{\rm max} \sim 0.006$. 
However, a lower limit cannot be set because the best-fit model cannot 
be distinguished from a point-source model within $2\sigma$.

\subsection{Binary-source Interpretation}

It is known that a subset of binary-source events can produce short-term 
anomalies similar to planetary perturbations and thus masquerade as planetary 
events \citep{Gaudi1998, Hwang2013, Hwang2017}.  We, therefore, check 
the possibility of the interpretation in which the perturbation is produced 
by a source companion.  The lensing magnification of a binary-source event 
corresponds to the flux-weighted mean of the magnifications  associated 
with the individual source stars, $A_1$ and $A_2$, i.e.,
\begin{equation}
A={A_1 F_1 + A_2 F_2 \over F_1+F_2 }=
{A_1+q_F A_2 \over 1+q_F}.
\end{equation} 
Here $q_F=F_2/F_1$ represents the ratio between the unmagnified fluxes 
of the individual source stars \citep{Han1998}.  We conduct modeling of 
the observed data with the binary-source interpretation.  For a binary-source 
event mimicking a planetary event, the flux ratio is usually very small and 
the faint source approaches close to the lens.  In this case, lensing 
magnifications during the perturbation can be affected by finite-source 
effects.  We, therefore, consider finite-source effects in the modeling.

In Figure~\ref{fig:four}, we compare the fits of the planetary and 
binary-source models in the neighborhood of the anomaly.  
The best-fit binary-source model yields a flux ratio $q_{F,I}\sim 0.005$.
One finds that the binary-source 
model yields an unsatisfactory description of the region before the major 
anomaly, i.e., $7878.5 \lesssim {\rm HJD}' \lesssim 7879.7$.  Numerically, 
we find that the binary-source model is worse than the planetary model by 
$\Delta\chi^2=175.2$.  We, therefore, exclude the binary-source interpretation.

\begin{figure}
\includegraphics[width=\columnwidth]{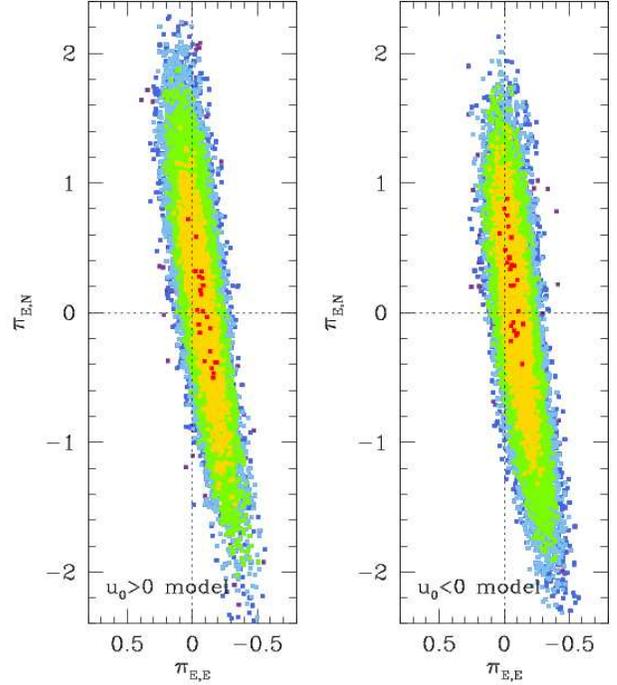}
\caption{
$\Delta\chi^2$ maps  in the $(\pi_{{\rm E},E},\pi_{{\rm E},N})$ plane. 
The left and right panels are for the $u_0>0$ and $u_0<0$ solutions, respectively. 
Color coding is same as in Fig.~\ref{fig:two} except that $n=1$.
The maps are based on the truncated data set, which is explained in 
Section 3.4.
}
\label{fig:five}
\end{figure}

\subsection{Higher-order Effects}

To precisely describe lensing light curves, it is often needed to consider 
higher-order effects.  In the case of OGLE-2017-BLG-0482, the event 
duration of $\Delta t=2 (1-u_0^2)^{1/2}t_{\rm E}\sim 80$ days, as measured 
by the time during which the source is within the Einstein ring, comprises 
an important portion of Earth's orbital period, i.e., 1 year.  In this case, 
the light curve can deviate from the one expected from a rectilinear 
lens-source relative motion due to the orbital motion of Earth: 
`microlens-parallax' effect \citep{Gould1992}.  Similarly, the orbital 
motion of the lens can also induce deviation in the lensing light curve: 
`lens-orbital' effect \citep{Albrow2000}.

Consideration of the higher-order effects requires additional parameters 
in lensing modeling. To account for the microlens-parallax effect, one 
needs two parameters of $\pi_{{\rm E},N}$ and $\pi_{{\rm E},E}$. 
These parameters denote the north and east components of the microlens-parallax 
vector, $\pivec_{\rm E}$, projected onto the sky along the north and east 
equatorial coordinates, respectively. The magnitude of the microlens 
parallax vector is
\begin{equation}
\pie=(\pi_{{\rm E},N}^2+\pi_{{\rm E},E}^2)^{1/2}={\pi_{\rm rel} \over \thetae},
\end{equation}
where $\pi_{\rm rel}={\rm au}(D_{\rm L}^{-1}-D_{\rm S}^{-1})$ is the lens-source 
relative parallax and $D_{\rm L}$ and $D_{\rm S}$ represent the distances to 
the lens and source, respectively. The direction of $\pivec_{\rm E}$ corresponds 
to the relative lens-source motion, $\muvec$. To the first-order approximation 
that the change rates of the binary separation and source trajectory angle are 
constant, the lens-orbital effect is described by two parameters, $ds/dt$ and 
$d\alpha/dt$, which represent the change rates of the binary separation and the 
source trajectory angle, respectively.  The measurement of the microlens parallax 
is important to determine the physical parameters of the lens mass and distance 
because $\pie$ is related to these parameters by
\begin{equation}
M={\thetae \over \kappa \pie},
\end{equation}
and
\begin{equation}
D_{\rm L}={\rm au \over \pie\thetae+\pi_{\rm S}},
\end{equation}
where $\kappa=4G/(c^2{\rm au})$ and $\pi_{\rm S}={\rm au}/D_{\rm S}$ is the 
parallax of the source.

In order to check the higher-order effects, we conduct additional modeling 
of the observed data.  From this modeling, we find that the fit substantially 
improves (by $\Delta\chi^2\sim 110$) with the consideration of the higher-order 
effects.  However, it is found find that the signal of the higher-order effects 
varies depending on the data sets.  From the inspection of the cumulative 
$\Delta\chi^2$ distributions as a function of time for the individual data sets, 
we find that most of the signal comes from the KMTC (by $\Delta\chi^2\sim 90$) 
and the KMTS  (by $\Delta\chi^2\sim 20$) data sets, while the signal from the 
other data sets is minor.  The inconsistency of the signal among different data 
sets are of concern because signals of the higher-order effects are subtle 
long-term deviations from the standard model, and thus they can be affected by 
the stability of photometry data.  We additionally check the reality of the signal 
by replacing the KMTC and KMTS photometry data with new ones processed using a 
different software, pySIS \citep{Albrow2000}.  From this, we find that the 
inconsistency between the KMTC+KMTS and the other data sets still persists.  
These results suggest the possibility that the KMTC and KMTS data are not 
stable enough to securely measure the higher-order parameters.

\begin{deluxetable}{lcc}
\tablecaption{Best-fit Lensing Parameters\label{table:one}}
\tablewidth{0pt}
\tabletypesize{\small}
\tablehead{
\multicolumn{1}{c}{Parameter} &
\multicolumn{2}{c}{Value}  \\
\multicolumn{1}{l}{} &
\multicolumn{1}{c}{$u_0>0$} &
\multicolumn{1}{c}{$u_0<0$}  
}
\startdata                                              
$\chi^2$                 & 1548.4                & 1548.6                \\
$t_0$ (HJD')             & $7873.948 \pm 0.013$  & $7873.952 \pm 0.013$  \\
$u_0$                    & $   0.059 \pm 0.002$  & $  -0.058 \pm 0.002$  \\
$t_{\rm E}$ (days)       & $  40.01  \pm 1.45 $  & $  40.39  \pm 1.21 $  \\ 
$s$                      & $   1.07  \pm 0.01 $  & $   1.07  \pm 0.01 $  \\
$q$ ($10^{-4}$)          & $   1.35  \pm 0.20 $  & $   1.41  \pm 0.30 $  \\
$\alpha$ (rad)           & $   0.369 \pm 0.023$  & $  -0.365 \pm 0.026$  \\
$\pi_{{\rm E},N}$        & $   0.19  \pm 0.75 $  & $  -0.17  \pm 0.73 $  \\
$\pi_{{\rm E},E}$        & $  -0.06  \pm 0.12 $  & $  -0.12  \pm 0.11 $  \\
$ds/dt$ (yr$^{-1}$)      & $  -1.94  \pm 0.82 $  & $  -1.96  \pm 0.96 $  \\
$d\alpha/dt$(yr$^{-1}$)  & $  -0.29  \pm 0.27 $  & $   0.87  \pm 0.50 $  \\  
$F_{s,{\rm OGLE}}$       & $  0.124  \pm 0.006$  & $  0.122  \pm 0.004$  \\
$F_{b,{\rm OGLE}}$       & $  0.029  \pm 0.005$  & $  0.031  \pm 0.004$         
\enddata
\tablecomments{ ${\rm HJD}'={\rm HJD}-2450000$.  }
\end{deluxetable}

Knowing the possibility of systematics in the KMTC and KMTS data, we mainly use 
the OGLE data set to measure the higher-order parameters.  The KMTNet survey 
started in 2015 season and thus the system is still under development.  On the 
other hand, the OGLE system is very stable from its 25-year operation since 1992.  
While the higher-order effect parameters are determined based on the overall shape 
of the lensing light curve, the planet parameters are determined by the planetary 
anomaly.  The overall light curve was well covered by the OGLE data, but the OGLE 
coverage around the planetary anomaly is poor.  In the analysis, we, therefore, 
use the combination of data sets with the whole OGLE data set plus partial data sets
from the other data sets.  The KMTNet+MOA data used in the analysis cover the anomaly 
region during $7875 < {\rm HJD}' < 7887$.  We exclude MOA data outside this 
region as well because of the instability in the baseline.

In Figure~\ref{fig:five}, we present the $\Delta\chi^2$ map of MCMC chains in 
the $(\pi_{{\rm  E},E},\pi_{{\rm E},N})$ plane obtained using the the restricted 
data set.  In the model 
considering the microlens-parallax effect, it is known that there exist a pair 
of degenerate solutions with $u_0>0$ and $u_0<0$ due to the mirror symmetry of 
the source trajectories between the two degenerate solutions \citep{Smith2003, Skowron2011}, 
and thus we check this so-called ecliptic degeneracy. We note that the lensing 
parameters of the two solutions resulting from the ecliptic degeneracy are 
approximately in the relation $(u_0, \alpha, \pi_{{\rm E},N}, d\alpha/dt)  
\leftrightarrow -(u_0, \alpha, \pi_{{\rm E},N}, d\alpha/dt)$.  From the 
distributions of MCMC points, we find that both $u_0>0$ and $u_0<0$ solutions 
result in similar values of $\pie$.  The measured 
microlens-parallax parameters are 
$(\pi_{{\rm E},N}, \pi_{{\rm E},E})=
(0.19 \pm 0.75, -0.06 \pm 0.12)$ for the $u_0>0$ model and 
$(-0.17  \pm 0.73, -0.12  \pm 0.11)$ for the $u_0<0$ model.  
It is found that the 
east component of the microlens-parallax vector, $\pi_{{\rm E},E}$, is well 
constrained but the uncertainty of the north component, $\pi_{{\rm E},N}$, 
is considerable.

\begin{figure}[t]
\epsscale{1.1}
\plotone{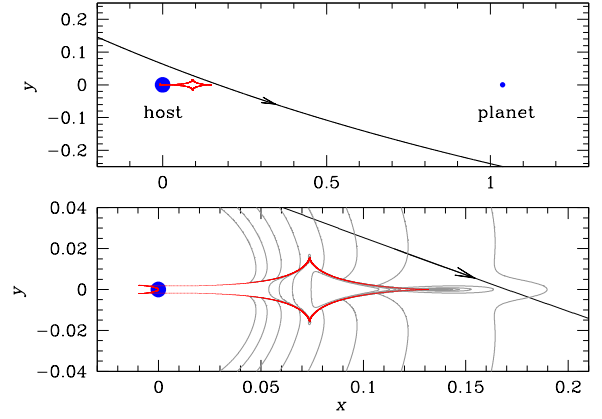}
\caption{
Geometry of the lens system.  The upper panel shows the positions of both 
the host and planet while the lower panel shows the zoom around the caustic.  
The curve with an arrow represents the source trajectory.  The lens components 
are marked by blue dots where the bigger one represents the host and the 
smaller one is the planet. The red cuspy closed curve is the caustic.  All 
lengths are scaled to the angular Einstein radius corresponding to the total 
mass of the lens.  The grey curves around the caustic represent the contours 
of lensing magnitications with $A=6$, 8, 10, 12, 14, 16, 18, and 20, respectively.  
}
\label{fig:six}
\end{figure}

In Table~\ref{table:one}, we list the lensing parameters of the best-fit 
solutions.  Since the ecliptic degeneracy is severe with $\Delta\chi^2<1$, 
we present both solutions with $u_0>0$ and $u_0<0$.  Also presented are the 
flux from the source $F_s$ and the blended light $F_b$ measured from the OGLE 
data.  We find that the event was produced by a planetary system with 
$s\sim 1.07$ and $q\sim 1.4\times 10^{-4}$.

In Figure~\ref{fig:six}, we 
present the lens system geometry (for the $u_0>0$ solution), in which the 
source trajectory with respect to the lens and caustic are shown.  Due to the 
proximity of the normalized planet-host separation to unity, i.e., $s\sim 1$, 
the caustics form a single closed curve for which the central and planetary 
caustics are merged into a single caustic, i.e., resonant caustic.  The 
planetary perturbation was produced when the source approached the strong 
planet-side cusp of the caustic located on the planet-host axis.  As mentioned 
in section 3.2, the source trajectory did not pass over the caustic and thus 
finite-source effects cannot be measured.  We note that the uncertainties of 
the lens-orbital parameters $ds/dt$ and $d\alpha/dt$ are big, indicating that 
the lens-orbital motion is poorly constrained.  Nevertheless it is important
to fit for these parameters because they can be correlated with the parallax 
parameters \citep{Batista2011,Skowron2011}.

\begin{figure}[t]
\includegraphics[width=\columnwidth]{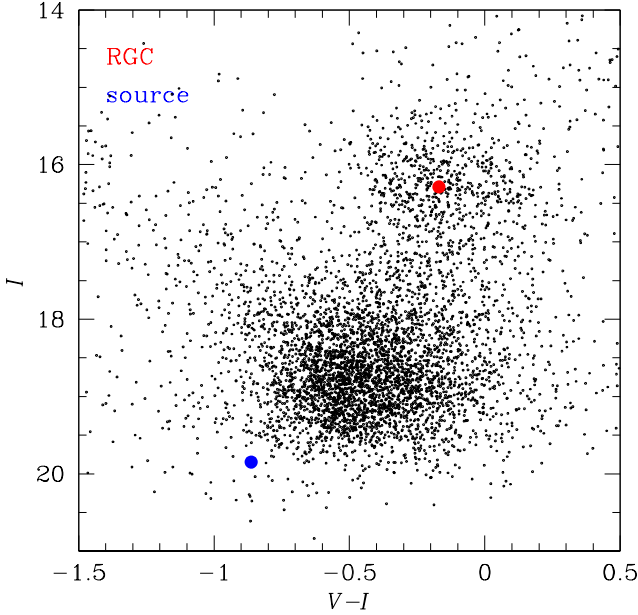}
\caption{
The locations of the source star (blue dot) and the centroid of red giant 
clump (RGC, red dot) in the instrumental color-magnitude diagram.  The source 
is unusually blue $(V-I)_0 = 0.37$ relative to most microlensed sources and 
therefore is inferred to lie in the Galactic disk rather than the bulge.
} 
\label{fig:seven}
\end{figure}

\subsection{Source Star}

In the analysis of lensing events, characterizing the source star is important 
because the angular Einstein radius, which is needed to determine the lens 
mass and distance by Equations (6) and (7), is derived from the angular source 
radius by $\thetae=\theta_*/\rho$. In the case of OGLE-2017-BLG-0482, the 
normalized source radius $\rho$ and thus the angular Einstein radius $\thetae$ 
cannot be measured.  However, the upper limit on $\rho$ leads to a lower limit 
on $\thetae$, and this lower limit may help to constrain the physical lens 
parameters. Thus determination of $\theta_*$ could be important depending 
on what limit is obtained.

We characterize the source based on its dereddened color $(V-I)_0$ and 
brightness $I_0$.  To determine $(V-I)_0$ and $I_0$ from the instrumental 
color $V-I$ and brightness $I$, we apply the method of \citet{Yoo2004} 
using the centroid of red giant clump (RGC) as a reference.  Figure~\ref{fig:seven} 
shows the locations of the source and the RGC centroid in the instrumental 
color-magnitude diagram.  With the offsets in color and brightness 
$\Delta(V-I,I)=(-0.69,3.56)$ with respect to the RGC centroid and the known 
dereddened color and brightness of the RGC centroid of 
$(V-I,I)_{0,{\rm RGC}}=(1.06,14.46)$ \citep{Bensby2013, Nataf2013}, it is 
estimated that the dereddened color and brightness of the source are 
$(V-I,I)_0=(V-I,I)_{0,{\rm RGC}}+\Delta(V-I,I)=(0.37,18.02)$.  We note that 
the relatively blue color indicates that the source is a main-sequence star 
located in the disk.  
Considering the faintness, the star is likely to 
lie behind the obscuring dust.

Once the dereddened $V-I$ color is measured, we then use the $VIK$ color-color 
relations of \citet{Bessell1988} to convert from $V-I$ to $V-K$ and apply 
the color/surface-brightness relation of \citet{Kervella2004} to obtain 
$\theta_* = 0.56\,\mu$as.  Combined with the $3\sigma$ upper limit we have 
derived on $\rho$ and the measured Einstein timescale, this implies
\begin{equation}
\thetae = {\theta_*\over\rho} > 0.093\,{\rm mas}
\end{equation}
and
\begin{equation}
\qquad
\mu = {\thetae\over t_{\rm E}} > 0.86 ~{\rm mas ~yr^{-1}} .
\end{equation}

We note that the lower limits of $\thetae$ and $\mu$ can be subject to additional 
uncertainties because of the uncertain
ratio between the extinction values towards the source and the RGC centroid.
The method using a color-magnitude diagram to derive $\theta_*$ assumes 
that the amount of extinction towards the source and RGC stars is same.  For 
OGLE-2017-BLG-0482, the source is likely to be in the disk, and thus the hypothesis 
of the same extinction may not be valid.  If the source suffers less extinction than 
RGC stars, it would be actually fainter and redder than if it were at the same 
distance as the RGC stars.  Therefore, the derived value of $\theta_*=0.56\,\mu$as 
is an upper limit and the additional uncertainty would propagate into 
$\theta_{\rm E, min}$ and $\mu_{\rm min}$.

\begin{figure}
\includegraphics[width=\columnwidth]{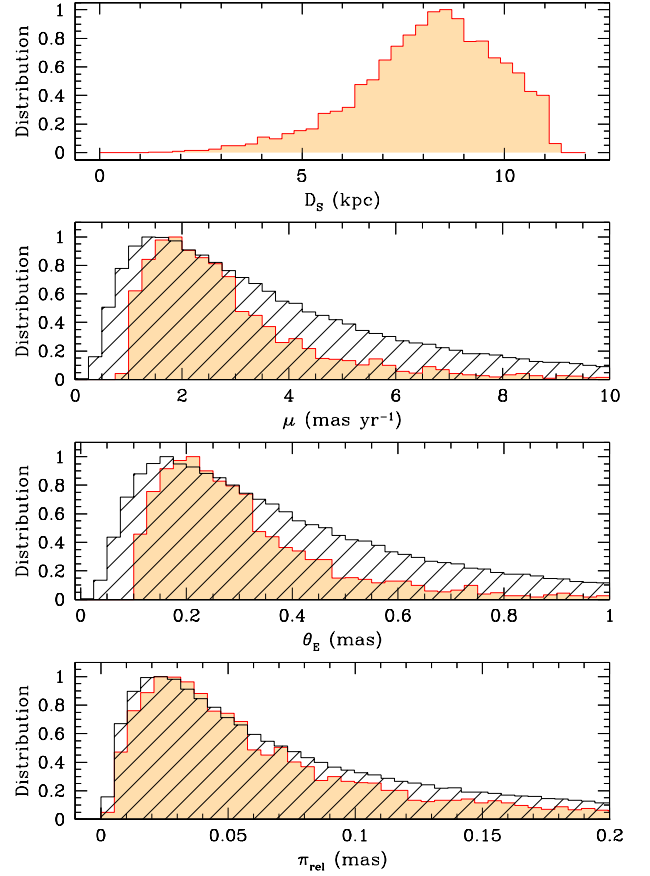}
\caption{
The distributions of 
the source distance ($D_{\rm S}$),
the angular Einstein radius $\thetae$ (middle panel), and the relative 
lens-source parallax $\pi_{\rm rel}={\rm au}(D_{\rm L}^{-1}-D_{\rm S}^{-1})$
produced from the Bayesian analysis using the adopted Galactic 
models. 
For $\mu$, $\thetae$, and $\pi_{\rm rel}$, 
we present two sets of distributions: one 
with (histogram filled with a yellow shade) and the other without (histogram 
shaded by slanted lines) the constraint of the measured $\te$ and $\pie$.
} 
\label{fig:eight}
\end{figure}

\begin{figure}
\includegraphics[width=\columnwidth]{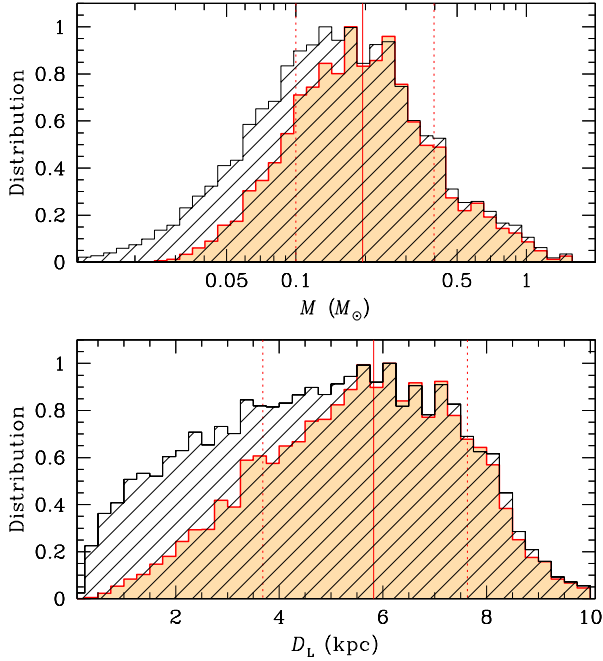}
\caption{Distributions of the mass (upper panel) and the distance to the lens 
(lower panel) obtained from the Bayesian analysis. The solid vertical line 
represents the median and the dotted lines represent the 1$\sigma$ range of 
the distribution.  In each panel, the histogram filled with a yellow shade 
represents the distribution including the constraint from the measured 
microlens-parallax parameters, while the histogram shaded by slanted lines 
represents the distribution without the parallax constraint.
} 
\label{fig:nine}
\end{figure}

\section{Physical Lens Parameters}

As described by Equations (6) and (7), if both $\theta_\e$ and $\pie$ are 
measured, then the mass $M$ and distance $D_{\rm L}$ can be uniquely 
determined.  In the present case, $\pie$ is measured but for $\thetae$ we 
have only a lower limit $\theta_{\rm E, min}$, which is given by Equation~(8).  
Therefore, we constrain the physical lens parameters by conducting 
a Bayesian analysis based on the constraints of the measured event timescale 
$\te$ and the microlens parallax $\pivec_{\rm E}$ combined with $\theta_{\rm E, min}$.

For the Bayesian analysis, one needs prior models of the mass function of 
lens objects and the density and dynamical distributions of Galactic matter.  
The source is a disk star and thus we need models to describe disk 
self-lensing events in which disk source stars are lensed by disk lenses.  
For the mass function, we employ \citet{Chabrier2003}.  
For the density distribution, we adopt the \citet{Han2003} model in which 
the disk is described by  a double-exponential disk.  For the velocity 
distribution, we use the dynamical model of \citet{Han1995}, in which the 
motion of disk objects is modeled by a Gaussian about the disk rotation speed.
OGLE-2017-BLG-0482 is likely to be a disk self-lensing event, where a disk star is 
lensed by a foreground disk star.
In the Bayesian analysis,   
we therefore locate both the source and lens following the disk matter distribution model. 
The top panel of Figure~\ref{fig:eight} shows the distribution $D_{\rm S}$.
The range of the source distance as measured by 1$\sigma$ uncertainty is
$D_{\rm S}=8.1_{-1.8}^{+1.6}$ kpc. 
We note that the presented $D_{\rm S}$ distribution is different from 
the distribution of disk stars because 
the lensing probability is higher for distant stars.

In the Bayesian analysis, we produce a large number of events by conducting 
a Monte Carlo simulation based on the prior models of the mass function, 
physical and dynamical distributions.  We estimate $M$ and $D_{\rm L}$ and 
their uncertainties from the distributions of events with timescales and 
microlens-parallax values located within the ranges of the measured values.  
In this process, we also impose the constraint of the lower limit of the 
angular Einstein radius, $\theta_{\rm E, min}$, given in Equation (8),
and the measured microlens parallax.
We impose the microlens-parallax constraint by computing the covariance 
matrix based on the MCMC chain in order to consider the 
uncertainties of the north and east components of $\pivec_{\rm E}$
and the orientation of the distributions in the $\pi_{{\rm E},E}$--$\pi_{{\rm E},N}$ plane.

In Figure~\ref{fig:eight}, we present the distributions of 
the relative lens-source proper motion $\mu$, the angular Einstein radius 
$\thetae$, and the relative lens-source parallax 
$\pi_{\rm rel}$
produced from the Bayesian analysis using the adopted Galactic models. 
We present two sets of distributions: one 
with (histogram filled with a yellow shade) and the other without (histogram 
shaded by slanted lines) the constraint of  
$\te$ and $\pie$.
From the comparison of the two sets of distributions, 
it is found that the constraint of $\te$ and $\pie$ on the
distributions is weak.
There exist several reasons for this. 
First, the source did not cross the
caustic and thus its relative size to the angular Einstein radius, $\rho$, cannot 
be constrained, preventing from measuring $\thetae=\theta_*/\rho$ 
and the relative proper motion between the source and the
lens, $\mu=\thetae/\te$.
Second, the upper limit given in Equation~(8) is not a strong constraint
because typical values of $\thetae$ are larger than $\sim 0.1$ mas.
Finally, the constraint on the microlens parallax $\pie$ is weak, as only one
projection, i.e., $\pi_{{\rm E},E}$, is constrained
but the other one, i.e., $\pi_{{\rm E},N}$, is poorly determined.

Figure~\ref{fig:nine} shows the distributions of the mass (upper panel) 
and the distance to the lens (lower panel) obtained from the Bayesian 
analysis.  In each panel, we present two distributions.  The histogram 
filled with a yellow shade represents the distribution obtained with the 
constraint of the event timescale and the microlens-parallax parameters, 
while the histogram shaded by slanted lines represents the distribution 
without the microlens-parallax constraint.  
From the comparison of the 
two distributions, it is found that the measured microlens parallax 
parameters enable to exclude 
lenses with very low masses located at small distances.  On the other hand, 
we find that the constraint of $\theta_{\rm E,min}$ on the physical lens 
parameters is weak to significantly impact the posterior distribution.

It is found that the masses of the planet and the host are
\begin{equation}
M_{\rm p}=9.0_{-4.5}^{+9.0}\ M_\oplus
\end{equation}
and
\begin{equation}
M_{\rm host}=0.20_{-0.10}^{+0.20}\ M_\odot ,
\end{equation}
respectively.  We note that the masses and their uncertainties are estimated 
as the median values and the standard deviations of the distributions obtained 
from the Bayesian analysis.  The mass of the planet is less than $10~M_\oplus$ 
and thus the planet is a super-earth according to the definition of 
\citet{Valencia2007}.  The mass of the host corresponds to that of late M-type 
dwarf.  Therefore, the lens is a planetary system composed of a super-earth 
and a low-mass M-dwarf host.

The lens is located at a distance from Earth of
\begin{equation}
D_{\rm L}=5.8_{-2.1}^{+1.8}\ {\rm kpc}.
\end{equation}
The projected separation of the planet from the host is 
\begin{equation}
a_\perp = 1.8_{-0.7}^{+0.6}\ {\rm au}.
\end{equation}

We note that the physical lens parameters corresponding to the $u_0<0$ solution
are similar to the presented values due to the similarity of the lensing parameters
between the two degenerate solutions except the sign of $\pi_{{\rm E},N}$.

\begin{figure}[t]
\includegraphics[width=\columnwidth]{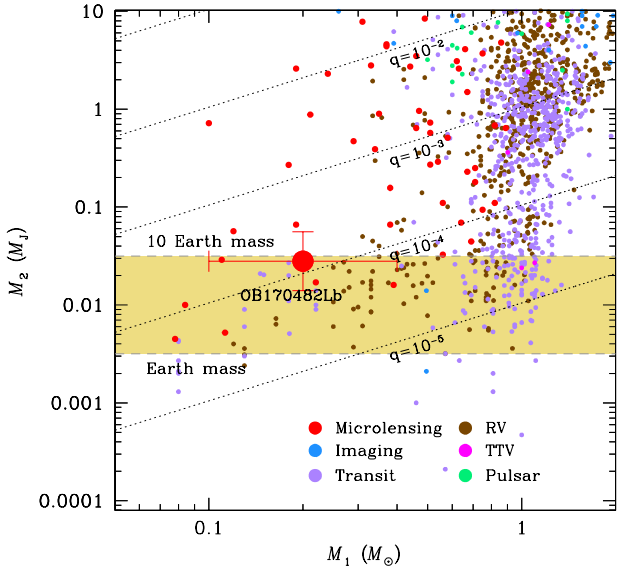}
\caption{Distributions of planets with known masses in the parameter space of 
the host mass $M_1$ and the planet mass $M_2$.  The position of OGLE-2017-BLG-0482L 
is marked by a big red dot.  The shaded area indicates the region of super-earths 
for which the best estimates of the planet masses
are in the range $1~M_\oplus < M_2 < 10~M_\oplus$.  The dotted 
lines represent the mass ratios.  The acronyms `RV' and `TTV' represent the 
radial-velocity and the transit-time-variation methods, respectively.
} 
\label{fig:ten}
\end{figure}

\begin{deluxetable*}{lllll}
\tablecaption{
Super-Earth microlensing planets with M-dwarf hosts \label{table:two}
}
\tablewidth{0pt}
\tablehead{
\multicolumn{1}{c}{Planet} &
\multicolumn{1}{c}{Planet mass}  &
\multicolumn{1}{c}{Host mass}  &
\multicolumn{1}{c}{Reference}  
}
\startdata                                              
OGLE-2005-BLG-390Lb  & $5.5_{-2.7}^{+5.5}\ M_\oplus$              & $0.22_{-0.11}^{+0.21}\ M_\odot$  &  \citet{Beaulieu2006}                       \\   
MOA-2007-BLG-192Lb   & $3.2_{-1.8}^{+5.2}\ M_\oplus$              & $0.084_{-0.012}^{+0.015}\ M_\odot$  &  \citet{Bennett2008}, \citet{Kubas2012}  \\
MOA-2009-BLG-266Lb   & $10.4\pm 1.7\ M_\oplus$                    & $0.56\pm 0.09\ M_\odot$  &  \citet{Muraki2011}                                 \\
MOA-2010-BLG-328Lb   & $9.2\pm 2.2\ M_\oplus$                     & $0.11\pm 0.01\ M_\odot$  &  \citet{Furusawa2013}                               \\
OGLE-2013-BLG-0341Lb & 2   $M_\oplus$         & 0.13 $M_\odot$    &  \citet{Gould2014}                                                             \\
OGLE-2016-BLG-1195Lb & $1.43_{-0.32}^{+0.45}\ M_\oplus$           & $0.078_{-0.012}^{+0.016} M_\odot$  &  \citet{Shvartzvald2017}                  \\
MOA-2012-BLG-505Lb   & $6.7_{-3.6}^{+10.7}\ M_\oplus$             & $0.10_{-0.05}^{+0.16}\ M_\odot$  &  \citet{Nagakane2017}                       \\
OGLE-2017-BLG-0482Lb & $9.0_{-4.5}^{+9.0}\ M_\oplus$              & $0.20_{-0.10}^{+0.20} M_\odot$  &  This paper
\enddata                                              
\end{deluxetable*}

\section{Discussion}

It is found that the planetary system OGLE-2017-BLG-0482L is composed of a 
super-earth orbiting a low-mass host star.  Due to the faintness of host stars, 
such planetary systems are difficult to detect using other major planet detection 
methods such as the radial-velocity (RV) and transit methods in which planets 
are indirectly found from observations of host stars.

To demonstrate the high efficiency of the microlensing method to these planetary 
systems, in Figure~\ref{fig:ten}, we present the distribution of planetary systems 
with known masses in the plane of the host mass $M_1$ and the planet 
mass $M_2$.  We mark the location of OGLE-2017-BLG-0482L by a big red dot.  In 
the plot, the shaded area indicates the region of super-earths where the 
planet masses are in the range $1~M_\oplus < M_2 < 10~M_\oplus$.  It shows 
that the microlensing method is sensitive to planets with low-mass hosts 
while the RV and transit methods are sensitive to planets 
orbiting solar-type stars.  
One also finds that the fraction of 
microlensing planets is especially high in the region of super-earths with 
very low-mass ($M_1 \lesssim 0.2~M_\odot$) hosts.  In Table~\ref{table:two}, 
we list the super-earth planets orbiting M-dwarf hosts detected using the microlensing 
method.

\section{Conclusion}

We analyzed the microlensing event OGLE-2017-BLG-0482, in which the light 
curve exhibited a short-term anomaly to the smooth lensing light curve.  
Analysis of the observed light curve indicated that the lens was a 
planetary system with a planet/host mass ratio of $q\sim 1.4\times 10^{-4}$. 
We measured the microlens parallax $\pivec_{\rm E}$ from the long-term deviation in 
the observed lensing light curve, but the angular Einstein radius $\thetae$ 
could not be measured.  Using the measured $t_{\rm E}$ and $\pivec_{\rm E}$, we found that the 
planetary system was composed of a super-earth and a late M-dwarf host.  The 
discovery of the planetary system demonstrates that microlensing provides 
an important tool to detect such planetary systems that are difficult to 
be detected by other methods.

\acknowledgments
Work by C.~Han was supported by the grant (2017R1A4A1015178) of
National Research Foundation of Korea.
The MOA project is supported by JSPS KAKENHI Grant Number JSPS24253004, JSPS26247023,
JSPS23340064, JSPS15H00781, and JP16H06287.
The OGLE project has received funding from the National Science Centre, Poland, grant 
MAESTRO 2014/14/A/ST9/00121 to A.~Udalski.  
Work by A.~Gould was supported by JPL grant 1500811.
Work by J.~C.~Yee was performed under contract with
the California Institute of Technology (Caltech)/Jet Propulsion
Laboratory (JPL) funded by NASA through the Sagan
Fellowship Program executed by the NASA Exoplanet Science Institute.
Work by YS was supported by an appointment to the NASA Postdoctoral Program at the Jet
Propulsion Laboratory, California Institute of Technology, administered by Universities Space
Research Association through a contract with NASA.
We acknowledge the high-speed internet service (KREONET)
provided by Korea Institute of Science and Technology Information (KISTI).
This research has made use of the KMTNet system operated by the Korea
Astronomy and Space Science Institute (KASI) and the data were obtained at
three host sites of CTIO in Chile, SAAO in South Africa, and SSO in
Australia.


\begin{thebibliography}{}
\bibitem[Alard \& Lupton(1998)]{Alard1998} Alard, C., \& Lupton, R.~H.\ 1998, \apj, 503, 325
\bibitem[Albrow et al.(2000)]{Albrow2000} Albrow, M.~D., Beaulieu, J.-P., Caldwell, J.~A.~R., et al.\ 2000, \apj, 534, 894
\bibitem[Albrow et al.(2009)]{Albrow2009} Albrow, M.~D., Horne, K., Bramich, D.~M., et al.\ 2009, \mnras, 397, 2009
\bibitem[Batista et al.(2011)]{Batista2011} Batista, V., Gould, A., Dieters, S., et al. \aap, 529, 102
\bibitem[Beaulieu et al.(2006)]{Beaulieu2006}  Beaulieu, J.-P., Bennett, D. P., Fouqu\'e, et al.\ P., 2006, Nature, 439, 437
\bibitem[Bensby(2013)]{Bensby2013} Bensby, T., Yee, J.~C., Feltzing, S., et al.\ 2013, \aap, 549, A147
\bibitem[Bennett \& Rhie(1996)]{Bennett1996} Bennett, D.~P., \& Rhie, S.~H.\ 1996, \apj, 472, 660
\bibitem[Bennett et al.(2008)]{Bennett2008}  Bennett, D.~P., Bond, I.~A., Udalski, A., et al.\ 2008, \apj, 684, 663
\bibitem[Bessell \& Brett(1988)]{Bessell1988} Bessell, M.~S., \& Brett, J.~M.\ 1988, \pasp, 100, 1134
\bibitem[Bond et al.(2001)]{Bond2001} Bond, I.~A., Abe, F., Dodd, R.~J., et al.\ 2001, \mnras, 327, 868
\bibitem[Bond et al.(2017)]{Bond2017} Bond, I.~A., Bennett, D.~P., Sumi, T., et al. 2017, \mnras, 469, 2434
\bibitem[Boss(1989)]{Boss1989} Boss, A.~P.\ 1989, \pasp, 101, 767
\bibitem[Chabrier(2003)]{Chabrier2003} Chabrier, G.\ 2003, \apj, 586, L133
\bibitem[Furusawa et al.(2013)]{Furusawa2013}  Furusawa, K., Udalski, A., Sumi, T., et al.\ 2013, \apj, 779, 91
\bibitem[Gaudi(1998)]{Gaudi1998} Gaudi, B.~S.\ 1998, \apj, 506, 533
\bibitem[Gaudi(2012)]{Gaudi2012} Gaudi, B.~S.\ 2012, \araa, 50, 411
\bibitem[Gould(1992)]{Gould1992} Gould, A.\ 1992, \apj, 392, 442
\bibitem[Gould \& Loeb(1992)]{GouldLoeb1992} Gould, A., \& Loeb, A.\ 1992, \apj, 396, 104
\bibitem[Gould et al.(2014)]{Gould2014}  Gould, A., Udalski, A., Shin, I.-G., et al.\ 2014, Science, 345, 46
\bibitem[Han(2006)]{Han2006} Han, C.\ 2006, \apj, 638, 1080
\bibitem[Han \& Gould(1995)]{Han1995} Han, C., \& Gould, A.\ 1995, \apj, 447, 53
\bibitem[Han \& Gould(2003)]{Han2003} Han, C., \& Gould, A.\ 2003, \apj, 592, 172
\bibitem[Han \& Jeong(1998)]{Han1998} Han, C., \& Jeong, Y.\ 1998, \mnras, 301, 231
\bibitem[Han et al.(2013)]{Han2013} Han, C., Jung, Y.~K., Udalski, A., er al.\ 2013\, \apj, 778, 38
\bibitem[Hwang et al.(2013)]{Hwang2013} Hwang, K.-H., Choi, J.-Y., Bond, I.~A., et al.\ 2013, \apj, 778, 55
\bibitem[Hwang et al.(2017)]{Hwang2017} Hwang, K.-H., Udalsk, A., Bond, I.~A., et al.\ 2017, arxiv:1711.09651
\bibitem[Kennedy et al.(2006)]{Kennedy2006} Kennedy, G.~M., Kenyon, S.~J., \& Bromley, B.~C.\ 2006, \apj, 650, L139
\bibitem[Kervella et al.(2004)]{Kervella2004} Kervella, P., Th\'evenin, F., Di Folco, E., \& S\'egransan, D.\ 2004, \aap, 426, 297
\bibitem[Kim et al.(2016)]{Kim2016}  Kim, S.-L., Lee, C.-U., Park, B.-G., et al.\ 2016, JKAS, 49, 37
\bibitem[Kim et al.(2018a)]{KimDJ2018} Kim, D.-J., Kim, H.-W., Hwang, K.-H., et al., 2018, \aj, 155, 76
\bibitem[Kim et al.(2018b)]{KimHW2018} Kim, H.-W., Hwang, K.-H., Kim, D.-J., et al. 2018, AAS, submitted
\bibitem[Kubas et al.(2012)]{Kubas2012} Kubas, D., Beaulieu, J.~P., Bennett, D.~P., et al.\ 2012,\ \aap, 540, 78 
\bibitem[Mao \& Paczy\'nski(1991)]{Mao1991} Mao, S., \& Paczy\'nski, B.\ 1991, \apj, 374, L37
\bibitem[Muraki et al.(2011)]{Muraki2011}  Muraki, Y., Han, C., Bennett, D.~P., et al.\ 2011, \apj, 741, 22
\bibitem[Nagakane et al.(2017)]{Nagakane2017} Nagakane, M., Sumi, T., Koshimoto, N., et al.\ 2017, \aj, 154, 35
\bibitem[Nataf et al.(2013)]{Nataf2013} Nataf, D.~M., Gould, A., Fouqu\'e, P., et al.\ 2013, \apj, 769, 88
\bibitem[Schechter et al.(1993)]{Schechter1993} Schechter, P.~L., Mateo, M., \& Saha, A.\ 1993, \pasp, 105, 1342
\bibitem[Shvartzvald et al.(2017)]{Shvartzvald2017}  Shvartzvald, Y., Yee, J. C., Calchi Novati, S., et al.\ 2017, \apj, 840, L3
\bibitem[Skowron et al.(2011)]{Skowron2011} Skowron,~J., Udalski,~A., Gould,~A., et al.\ 2011, \apj, 738, 87
\bibitem[Smith et al.(2003)]{Smith2003} Smith, M.~C., Mao, S., \& Paczy\'nski, B,\ 2003, \mnras, 339, 925
\bibitem[Sumi et al.(2003)]{Sumi2003}  Sumi, T., Abe, F., Bond, I.~A., et al.\ 2003, \apj, 591, 204
\bibitem[Udalski(2003)]{Udalski2003} Udalski, A.\ 2003, Acta Astron., 53, 291
\bibitem[Udalski et al.(2015)]{Udalski2015} Udalski,~A., Szyma\'nski,~M.~K., \& Szyma\'nski,~G.\ 2015, Acta Astron., 65, 1
\bibitem[Valencia et al.(2007)]{Valencia2007} Valencia, D., Sasselov, D.~D., \& O'Connell, R.~J.\ 2007, \apj, 656, 545
\bibitem[Wo\'zniak(2000)]{Wozniak2000} Wo\'zniak, P.~R.\ 2000, Acta Astron., 50, 421
\bibitem[Yee et al.(2012)]{Yee2012} Yee, J.~C., Shvartzvald, Y., Gal-Yam, A., et al.\ 2012, \apj, 755, 102
\bibitem[Yoo et al.(2004)]{Yoo2004} Yoo, J., DePoy, D.~L., Gal-Yam, A., et al.\ 2004, \apj, 603, 139
\end{thebibliography}
\end{document}